# 3D metrology with a laser tracker inside a vacuum chamber for NISP test campaign

Anne Costille[*,a], Florent Beaumont[b], Eric Prieto[a], Michael Carle[a], Romain Pawlowski[c], Thierry Roux[c], Olivier Dupuy[c], Christophe Fabron[a]
[a]Aix Marseille Univ, CNRS, CNES, LAM UMR 7326, 13388, Marseille, France, [b] DAES SA ITER organisation, 13067 St Paul Lez Durance Cedex – France, [c]Symétrie, Nîmes, France

## ABSTRACT

In the frame of the test of NISP instrument for ESA Euclid mission, the question was raised to perform a metrology measurement of different components during the thermal vacuum test of NISP instrument. NISP will be tested at Laboratoire d'Astrophysique de Marseille (LAM) in ERIOS chamber under vacuum and thermal conditions in order to qualify the instrument in its operating environment and to perform the final acceptance test before delivery to the payload. One of the main objectives of the test campaign will be the measurement of the focus position of NISP image plane with respect to the EUCLID object plane. To simulate the EUCLID object plane, a telescope simulator with a very well know focal distance will be installed in front of NISP into ERIOS chamber. We need to measure at cold and vacuum the position of reflectors installed on NISP instrument and the telescope simulator. From these measurements, we will provide at operational temperature the measurement of references frames set on the telescope simulator and NISP, the knowledge of the coordinates of the object point source provided by the telescope simulator and the measurement of the angle between the telescope simulator optical axis and NISP optical axis. In this context, we have developed a metrology method based on the use of a laser tracker to measure the position of the reflectors inside ERIOS. The laser tracker is installed outside the vacuum chamber and measure through a curved window reflectors put inside the chamber either at ambient pressure or vacuum pressure. Several tests campaigns have been done at LAM to demonstrate the measurement performance with this configuration. Using a well know reflectors configuration, we show that it is possible to correct the laser tracker measurement from the window disturbances and from the vacuum impact. A corrective term is applied to the data and allows retrieving the real coordinates of the reflectors with a bias lower than 30µm, which is lower than the laser tracker measurement uncertainties estimated at 60µm. No additional error term of the laser tracker measurement is observed when using the laser tracker with the curved window and in vacuum, comparing with a classical use of the laser tracker. With these test campaign, we have been able to demonstrate the possibility to use a laser tracker to measure in real time during a vacuum thermal test the position of different mechanical parts into a vacuum chamber with an accuracy better than 60µm.

## 1. INTRODUCTION

EUCLID mission[1] has been selected by ESA in 2012 in the context of the Cosmic Vision program to study the nature of the dark energy and the dark matter. The mission is designed to map the geometry of the dark Universe by investigating the distance-redshift relationship and the evolution of cosmic structures thanks to two scientific instruments: the Near Infrared Spectroscopic Photometer (NISP)[2] and the Visible Instrument (VIS)[3]. The laboratoire d'Astrophysique de Marseille (LAM) is deeply involved into the development of the NISP instrument as it is responsible for the integration and test of the instrument. NISP is dedicated to measure the redshift of millions of galaxies and to analyze their spatial distribution in the Universe thanks to two instrumental modes: the photometry mode uses broadband filters to acquire image of the galaxies, the spectroscopic mode used grisms to obtain there spectra. The broadband filters and grisms are mounted on two rotating wheels allowing to switch easily from one mode to another and to acquire data of the same field of view of space.

*anne.costille@lam.fr; phone +33491055978; fax; +33491621190, www.lam.fr

NISP instrument is made by an European consortium led by Centre National des Etudes Spatiales (CNES) that includes laboratory and industries mainly from France, Germany, Italy and Spain. It will be fully assembled and aligned at LAM in France. Then it will be tested under vacuum and thermal conditions in ERIOS chamber, a 90m3 chamber developed by LAM to test optical instruments at cryogenics temperature and high vacuum. NISP Proto Flight Model (PFM) will be tested during two Thermal Balance/Thermal Vacuum (TB/TV) in ERIOS, the first test being scheduled by end of 2018. The goal of these TB/TV is to qualify the instrument in its operating environment before delivery to the payload.

One of the main objectives of the test campaign[4] will be the measurement of the focus position of NISP image plane with respect to the EUCLID object plane to ensure a good focalisation of NISP instrument on EUCLID telescope. It has been identified very early in the project the need to develop a metrology mean to measure at cold during the TB/TV the relative position between NISP instrument and a simulator of EUCLID telescope. A special Metrology Verification System (MVS)[4,5] has been designed and studied for NISP test in ERIOS to measure during the TB/TV test the position of a set of reflectors set on the EUCLID telescope simulator and NISP instrument itself. The MVS concept is based on the use of a laser tracker, installed outside the vacuum chamber that measures reflectors installed inside the vacuum chamber through a curved window. A preliminary study, described in [5] has shown the feasibility of the measurement with a laser tracker through a curved window. But many questions were raised by this test and in particular what is the impact of the measurement configuration onto the uncertainties budget of the laser tracker. A large test campaign has been done in 2017 to study the measurement performance with the MVS configuration. We present in this article the results of this test campaign. First we describe the goal of the metrology in NISP context and the verification plan we have elaborated to validate the performance of the laser tracker. We present the mechanical standard specially designed to test the metrology in ERIOS, which has been fully designed and tested by Symétrie company[6]. We compare the measurement accuracy reached with a Coordinates Measurement Machine (CMM) and the laser tracker. Then we study the measurement accuracy obtained with the laser tracker when we measure through the curved window in ambient pressure and vacuum pressure. For both configurations, we propose a corrective term to correct the data from the disturbances introduced by the window and the vacuum. Finally we conclude on the feasibility of the metrology measurement for NISP test campaign.

## 2. METROLOGY CONFIGURATION FOR NISP TEST CAMPAIGN

### 2.1 Goal of NISP metrology measurement

NISP instrument will be tested in cold and vacuum by end of 2018 to validate its performance. In particular, it is required to perform a metrology during the cold test of NISP instrument to verify the instrument optical interfaces with the PLM at cold. In NISP context, the optical interfaces are obtained with the measurement of the object plane localization R_nisp and the measurement of the optical axis position with respect to a well-defined reference system set on NISP instrument and called R_p1. The position of the optical axis is measured during the NISP integration campaign thanks to an accurate Coordinate Measurement Machine (CMM). The localization of the object plane needs to be measured during the NISP TB/TV. The principle of the measurement is the following and it is shown in Figure 1:

- A Telescope Simulator (NI-TS) simulates EUCLID telescope and mimics the EUCLID object plane i.e. the optical interface between NISP and EUCLID. A set of reflectors is installed on the backside of the NI-TS to form a reference system called R_ts that can be measured at cold with the laser tracker. The measurement of the R_ts provides the position of the focus object of NI-TS called F_ts in R_ts and then indicates the object plane localization. The relationship between R_ts and F_ts is measured prior to NISP TB/TV with a CMM: $R_{ts} = M_{ts} x\ F\_ts$. When we move the NI-TS in NISP Field of View (FoV), we can then mimic R_nisp localization.

- During NISP TB/TV test, the NI-TS is focused on NISP instrument thanks to a through focus method. When the best focus position of the telescope is found, the laser tracker measures the position of the reflectors set on NI-TS, to know the position of F_ts, and on NISP instrument to calculate R_p1. The NI-TS is moved in the FoV to simulate R_nisp and then we measure the best focus position between R_p1 and R_nisp. This measurement needs to be done at Operational Temperature (OT) as NISP detectors cannot be used at warm temperature.

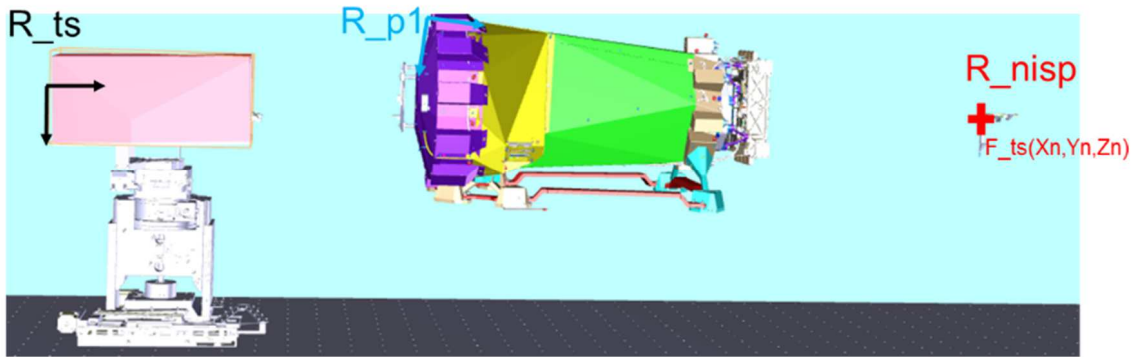

*Figure 1. Schematic of the measurement of the reference system on NISP (R_p1), on NI-TS (R_ts) with the localisation of the theoretical R_nisp.*

### 2.2 MVS description and verification plan

The MVS is a set of metrological means developed for NISP TB/TV test[4]. It is made of a AT402 laser tracker installed outside ERIOS chamber in front of a curved window. The laser tracker measures reflectors installed on NISP instrument and NI-TS in an environment at vacuum (P < 1.3 10e-6 mbar) and cold temperature (T < 130K). The MVS configuration is shown in Figure 2. The laser tracker is a well-known instrument for measurement in room temperature in a "normal" environment. In such conditions, AT402 measurement uncertainty is defined as the deviation between a measured coordinate and the nominal coordinate of that point and depends of the distance between the laser tracker and the measured point. AT402 uncertainty is estimated at +/-15µm + 6µm/m. Repeatability is estimated at 5µm and accuracy at 10µm.

As indicated by Leica, the laser tracker is not compatible with vacuum and cold temperature. This explains the configuration chosen for NISP TB/TV test. We have then to demonstrate the feasibility of the measurement through a curved window and the uncertainties of the measurements obtained in such configuration. Preliminary tests[5] have shown that the main contributors to the laser tracker errors in our configuration are:

- The window, as the laser tracker does not "know" that there is a window between the reflector and itself;
- The vacuum, as the laser tracker does not "know" that the index is different into the vacuum chamber.

The contribution from the cold environment is supposed to be negligible compared to the two main contributors to the errors. We have decided to elaborate a verification plan of the MVS based on the analysis of the impact of the window and the vacuum on the measurement. Adding a test at cold was too restrictive, too long and too expensive for the metrology verification plan. The verification plan strategy is based on the use of a mechanical standard representative of the NISP TB/TV reflectors configuration inside ERIOS. The mechanical standard represents in term of geometry and distance the reference systems R_ts and R_p1 that we will measure during NISP TB/TV test:

- 5 reflectors mimic the NI-TS reference system;
- 5 reflectors mimic NISP reference system;
- 5 reflectors mimic the R_nisp localization;
- Additional reflectors are installed onto the mechanical standard to probe a large and representative volume.

The design of the mechanical standard is shown in Figure 3 and has been done by Symétrie company [6]. The mechanical standard made in INVAR shall be well known and stable in time to ensure that we "know" what we are measuring. It has been measured precisely under a CMM (with an accuracy of (+/-1.9 + L/300)µm) so that the knowledge of the reflectors positions is very accurate. The mechanical standard is then consider as a reference or a caliber for the verification plan of the MVS. The main specification of the frame is to ensure that no movement of the reflectors are present between ambient and vacuum pressure at ambient temperature (+/- 2°C).

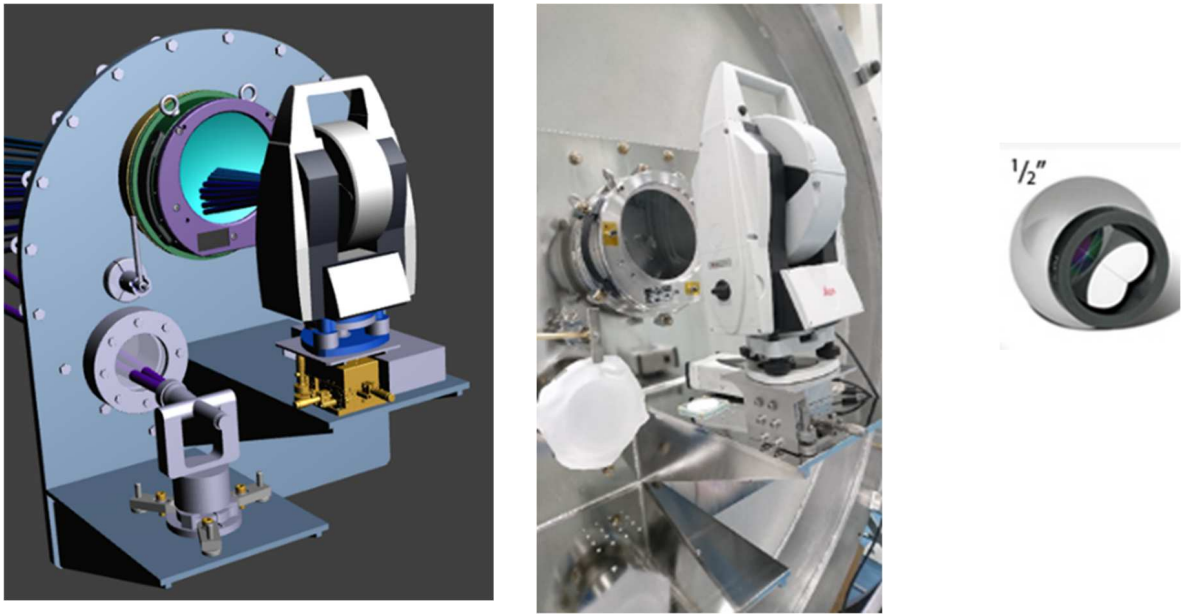

*Figure 2. From left to right: CAD view of the laser tracker in front of the curved window, picture of the set-up on ERIOS chamber, 0.5'' inch reflector type used for NISP TB/TV test.*

The verification plan is based on a sequential analysis:
- Accurate measurement of the mechanical standard with a CMM. The design, analysis and the measurement of the mechanical standard has been subcontracted to Symétrie company [6].
- Measurement of the mechanical standard with a laser tracker with no window to estimate the laser tracker uncertainties.
- Measurement of the mechanical standard with a laser tracker through the curved window to estimate the error introduced by the window. Alignment procedure of the laser tracker in front of the window is defined during this step that has been done in collaboration with Symétrie company.
- Measurement of the mechanical standard with the laser tracker through the curved window in vacuum and installed inside ERIOS chamber. The goal is to estimate the error introduced by the vacuum and ERIOS conditions (vibrations environment). The test configuration is shown in Figure 4.

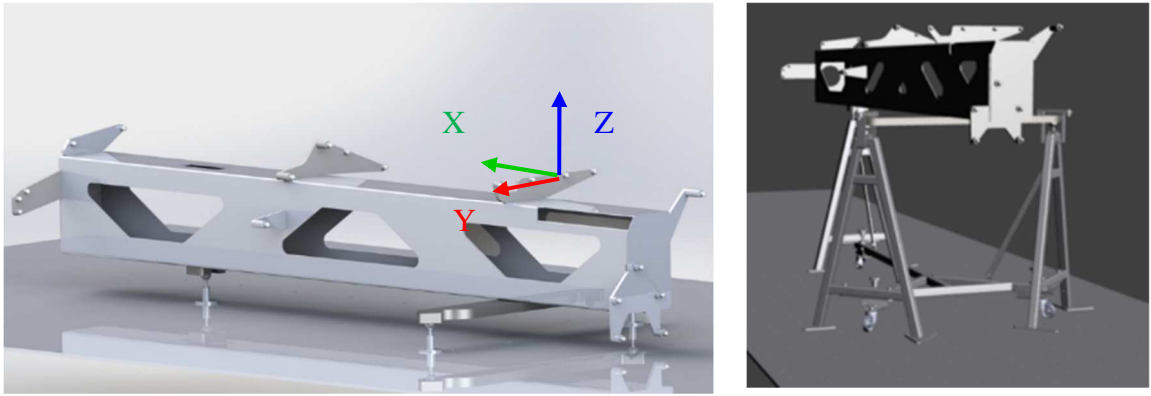

*Figure 3. CAD view of the mechanical standard without its feet (left) and on its mechanical interface (right). The reference system R_et of the mechanical standard is shown.*

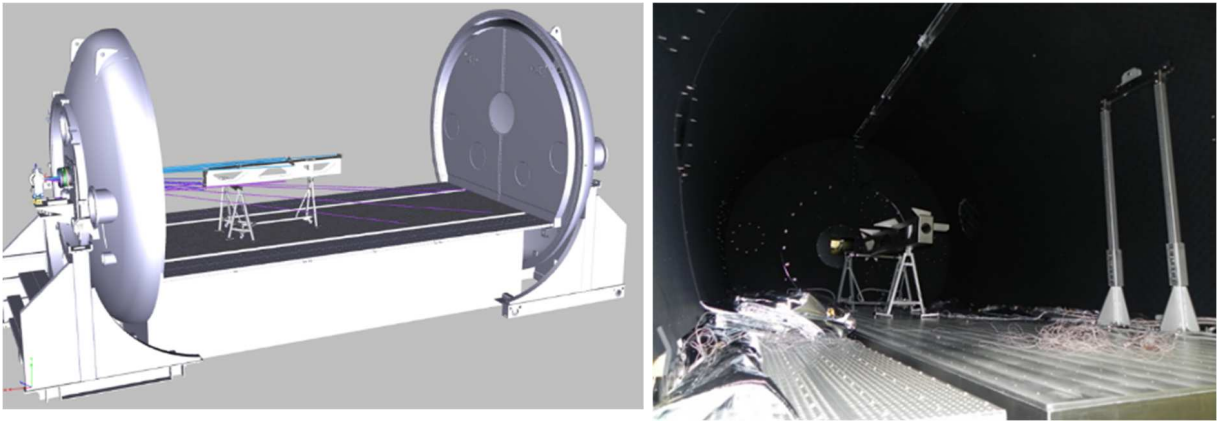

*Figure 4. 3D CAD model of the mechanical standard test in ERIOS with the laser tracker (left) and picture of the configuration inside ERIOS for the test (right).*

## 3. COMPARISON BETWEEN CMM AND LASER TRACKER MEASUREMENTS

### 3.1 Measurement of the mechanical standard reference values

The first step of the verification plan consists in the knowledge of the reflector positions set on the mechanical standard. As indicated before, the design and the full characterization and analysis of the mechanical standard has been done by the company Symétrie. The mechanical standard has been measured with an accurate CMM own by Symétrie to define with an accuracy of 2µm the localization of the 20 reflectors expressed in a reference system attached to the mechanical standard called R_et thanks to 3 reflectors and shown in Figure 3. The coordinates of the reflectors, called hereafter $(x_i, y_i, z_i)_0$, are considered as a reference for all measurements done during the verification campaign. All measurements are compared to this set of reference coordinates.

### 3.2 Measurement of the mechanical standard with a laser tracker

The second step of our verification plan is the measurement of the mechanical standard with the laser tracker, without a window, to compare the accuracy reached with the laser tracker with the CMM measurement. The analysis done by Symétrie shows that the accuracy of the laser tracker is estimated to 60µm. This budget takes into account the repeatability of the mechanical standard measurement in its environment, shown in Figure 5-left, the uncertainties of the laser tracker indicated by Leica (15µm + 6µm/m), all uncertainties linked to the evolution of the mechanical standard with temperature, etc and the uncertainties linked to the reference system measurement. It is interesting to note that the accuracy on X direction (direction of the laser measurement) is much more precise than on Y and Z directions. It is due to the fact that the measurement lays on the distancemeter of the laser tracker which is very accurate. The Y and Z direction measurements are less accurate due to the rotation of the laser tracker head and the accuracy of its motors. On Figure 5-right, we present the difference between the coordinates measured with the laser tracker and the coordinates measured with the CMM to compare both instruments. We can see that the absolute difference is not larger than 25µm: the laser tracker is quite a very accurate machine to measure an object in 3D. In the rest of the document, we compare the results obtained with the laser tracker with the value of 60µm. We consider that a result lower than 60µm is below the measurement error and then is considered as correct with no particular bias.

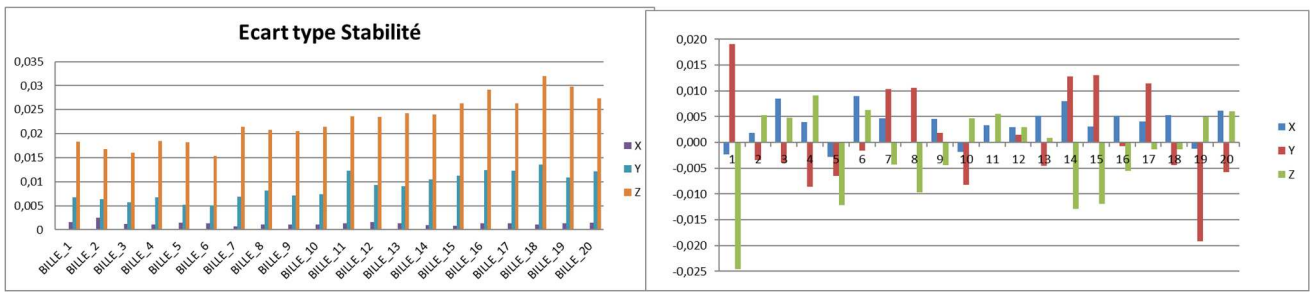

*Figure 5. Left: Measurement of the repeatability of the laser tracker on X, Y and Z coordinates for 20 reflectors. The values obtained are the standard deviation obtained with 20 measurements done during 10hours. Right: Difference of the X, Y and Z coordinates of 20 reflectors obtained with the laser tracker and with the CMM. Y-axis are given in mm, X-axis are the number of the reflectors.*

## 4. MEASUREMENT WITH A LASER TRACKER THROUGH A WINDOW

### 4.1 Measurement principle and corrective factor

The principle of the measurement of the laser tracker through a curved window is described in [5]. The concept is based on the alignment of the laser tracker on the center of curvature of the window. In this configuration, the measurement done by the laser tracker is disrupted by the window and then the X, Y and Z coordinates measured by the laser tracker are wrong. If we look at the spherical coordinates, we can see that only the distance ρ is disrupted by the window contribution and that the error on the angles θ and φ are negligible if we are aligned. According to the formula indicated in [5], when we measure through a curved window with the same index on both sides of the window the error on the distance measurement should be of $e(1 - n_v)$ where $e$ is the window thickness (29.798mm in our case) and $n_v$ the glass index (1.455 in our case). We have to put a corrective factor equal to 13.5mm for our configuration.

To validate this number, we have made measurements with Symétrie in a dedicated configuration: the laser tracker and the mechanical standard have been installed in a configuration similar to the final configuration inside ERIOS. First the mechanical standard is measured with the laser tracker and no window, the coordinates $(\rho_i, \theta_i, \varphi_i)_1$ of each reflector are obtained. Then the window is installed between the laser tracker and the mechanical standard so that the laser tracker is put on the center of curvature of the window. The same set of reflectors is measured and coordinates $(\rho_i, \theta_i, \varphi_i)_2$ are obtained. We then compare the coordinates $(\rho_i, \theta_i, \varphi_i)_1$ with $(\rho_i, \theta_i, \varphi_i)_2$ to see the difference. Results are shown in Table 1-left. From this analysis, one can see that the error on the distance ρ is almost constant and equal to 13.93mm and the error on the angles (θ, ρ) are small and also constant. Then we have applied the corrective factor found to obtain corrected coordinates $(\rho_i, \theta_i, \varphi_i)_3$ and compare them with the coordinates without the window $(\rho_i, \theta_i, \varphi_i)_1$. Results are shown in Table 1-right. We can see that once corrected the error are far below the measurement error of the laser tracker and then the correction applied is correct. Correction of the angles does not add a real improvement of the values obtained, it is then not necessary to apply this correction as the main contributor to the error is the distance error. We can note that the corrective factor is larger than the one we have calculated theoretically. This is linked to the uncertainties of the index of the glass, which is not perfectly known. For the rest of our study, we have decided to rely on the results obtained experimentally. During the test with Symétrie, we have concluded that the measurement obtained through the window shall be corrected of the following values:

- $\rho_{corr} = \rho_{mes} - 13.93$ mm
- θ and φ are not corrected. Errors on their values are negligible.

For all data measured with the window, we have decided to apply the corrective scheme shown in Figure 6. In addition, we study in the rest of this document the coordinates of the reflectors in the reference system of the mechanical standard R_et and compare them with the values obtained with the CMM $(x_i, y_i, z_i)_0$ for each reflector. We use the criterion called the correction error to analyse the data. It consists of the calculation of the mean value of the difference between the coordinates measured (with or without correction) and the reference values from the CMM for all reflectors.

*Table 1. Left: Difference of the coordinates measured with the window and without the window. Right: difference of coordinates measured with a window and corrected of ρ or ρ, θ, φ and without the window.*

$(\rho_i, \theta_i, \varphi_i)_2 - (\rho_i, \theta_i, \varphi_i)_1$

| Bille # | Δρ (mm) | Δθ (°) | Δφ (°) |
|---|---|---|---|
| BILLE 1 | 13.929 | -0.005 | -0.009 |
| BILLE 2 | 13.925 | 0.005 | 0.010 |
| BILLE 3 | 13.926 | -0.005 | -0.009 |
| BILLE 4 | 13.927 | -0.005 | -0.009 |
| BILLE 5 | 13.926 | -0.005 | -0.009 |
| BILLE 6 | 13.925 | 0.005 | 0.009 |
| BILLE 7 | 13.930 | -0.005 | -0.009 |
| BILLE 8 | 13.930 | -0.005 | -0.009 |
| BILLE 9 | 13.928 | -0.005 | -0.009 |
| BILLE 10 | 13.928 | -0.005 | -0.009 |
| BILLE 11 | 13.927 | -0.005 | -0.009 |
| BILLE 12 | 13.930 | -0.005 | -0.009 |
| BILLE 13 | 13.929 | 0.005 | 0.009 |
| BILLE 14 | 13.931 | -0.005 | -0.009 |
| BILLE 15 | 13.934 | -0.005 | -0.009 |
| BILLE 16 | 13.940 | -0.005 | -0.009 |
| BILLE 17 | 13.940 | -0.004 | -0.009 |
| BILLE 18 | 13.933 | -0.005 | -0.009 |
| BILLE 19 | 13.931 | 0.004 | 0.009 |
| BILLE 20 | 13.935 | -0.004 | -0.009 |
| Min | 13.924 | -0.005 | -0.010 |
| Max | 13.940 | -0.004 | -0.009 |
| Moy | 13.930 | -0.005 | -0.009 |
| Ecart Type | 0.005 | 0.000 | 0.000 |

$(x_i, y_i, z_i)_3 - (x_i, y_i, z_i)_1$

| Bille # | Correction ρ | | | Correction ρ θ φ | | |
|---|---|---|---|---|---|---|
| | ΔX (mm) | ΔY (mm) | ΔZ (mm) | ΔX (mm) | ΔY (mm) | ΔZ (mm) |
| BILLE 1 | 0.015 | 0.034 | -0.005 | 0.015 | 0.029 | -0.006 |
| BILLE 2 | 0.016 | -0.003 | 0.031 | 0.016 | -0.007 | 0.030 |
| BILLE 3 | 0.011 | 0.023 | 0.015 | 0.011 | 0.021 | 0.016 |
| BILLE 4 | 0.012 | 0.014 | 0.017 | 0.012 | 0.013 | 0.017 |
| BILLE 5 | 0.010 | 0.025 | 0.006 | 0.010 | 0.025 | 0.005 |
| BILLE 6 | 0.005 | 0.013 | -0.012 | 0.005 | 0.013 | -0.012 |
| BILLE 7 | 0.004 | 0.019 | 0.013 | 0.004 | 0.019 | 0.014 |
| BILLE 8 | 0.000 | 0.000 | 0.000 | 0.000 | 0.000 | 0.000 |
| BILLE 9 | 0.004 | -0.015 | 0.030 | 0.004 | -0.015 | 0.031 |
| BILLE 10 | 0.000 | 0.011 | 0.000 | 0.000 | 0.011 | 0.000 |
| BILLE 11 | 0.003 | 0.021 | 0.023 | 0.003 | 0.023 | 0.022 |
| BILLE 12 | 0.001 | 0.000 | 0.014 | 0.001 | -0.001 | 0.014 |
| BILLE 13 | 0.004 | 0.008 | 0.006 | 0.004 | 0.008 | 0.006 |
| BILLE 14 | -0.003 | -0.008 | -0.016 | -0.004 | -0.008 | -0.013 |
| BILLE 15 | 0.003 | -0.015 | -0.017 | 0.003 | -0.014 | -0.008 |
| BILLE 16 | 0.000 | 0.000 | 0.000 | 0.000 | 0.000 | 0.000 |
| BILLE 17 | -0.001 | 0.035 | 0.015 | -0.001 | 0.034 | 0.014 |
| BILLE 18 | 0.002 | 0.028 | 0.023 | 0.002 | 0.026 | 0.023 |
| BILLE 19 | 0.008 | 0.011 | 0.040 | 0.008 | 0.013 | 0.040 |
| BILLE 20 | 0.003 | 0.023 | 0.025 | 0.003 | 0.021 | 0.025 |
| Min | -0.003 | -0.025 | -0.017 | -0.004 | -0.015 | -0.013 |
| Max | 0.016 | 0.035 | 0.040 | 0.016 | 0.034 | 0.040 |
| Moy | 0.005 | 0.012 | 0.008 | 0.005 | 0.015 | 0.008 |
| Ecart Type | 0.006 | 0.017 | 0.016 | 0.006 | 0.017 | 0.016 |

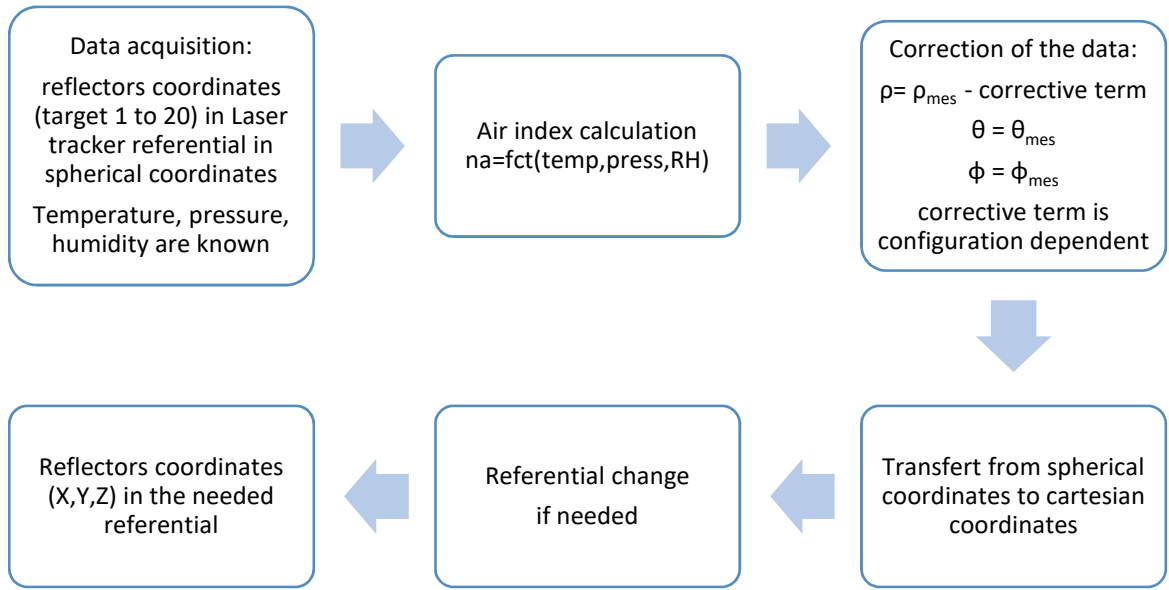

*Figure 6. Data reduction scheme for measurement with a laser tracker through a window.*

### 4.2 Measurement error with the window

After the test done at Symétrie, the mechanical standard, the window and the laser tracker are installed on ERIOS chamber. The laser tracker is aligned on the center of curvature of the window and the coordinates of the reflectors are measured with the laser tracker and then corrected with the corrective factor described in subsection 4.1. We analyze the measurement error due to the window i.e. the difference between the real value of the reflector position in R_et, measured with the CMM and called the reference values $(x_i, y_i, z_i)_0$, and the values of the reflector position obtained after correction of the data. Figure 7 presents the mean error values of the coordinates over 900 measurements (42 hours with a measurement done every minute). Figure 8 shows the difference for each reflector from the reference values. For both cases, we study the values without correction of the data and with the correction data applied. We can see that the correction of the data allows having a residual error with respect to the reference lower than the laser tracker uncertainties: we are really correcting the data and the residual error is included between 20 and 40µm, which is lower than the laser tracker uncertainties for the test configuration considered. This error varies with the time ant the measurement uncertainties have also an impact on the value itself. From this analysis, we can conclude that the correction of the data is consistent and that the measurement error after correction is lower than the laser tracker uncertainties. Figure 8 shows the mean error on the X, Y and Z coordinates of the reflectors over the 900 measurements acquired. We can see that the correction of the data is important and allows obtaining the coordinates of the reflectors within the laser tracker measurement uncertainties. The corrective factor applied is correct and constant in time as these measurements have been done several months after the test done at Symétrie.

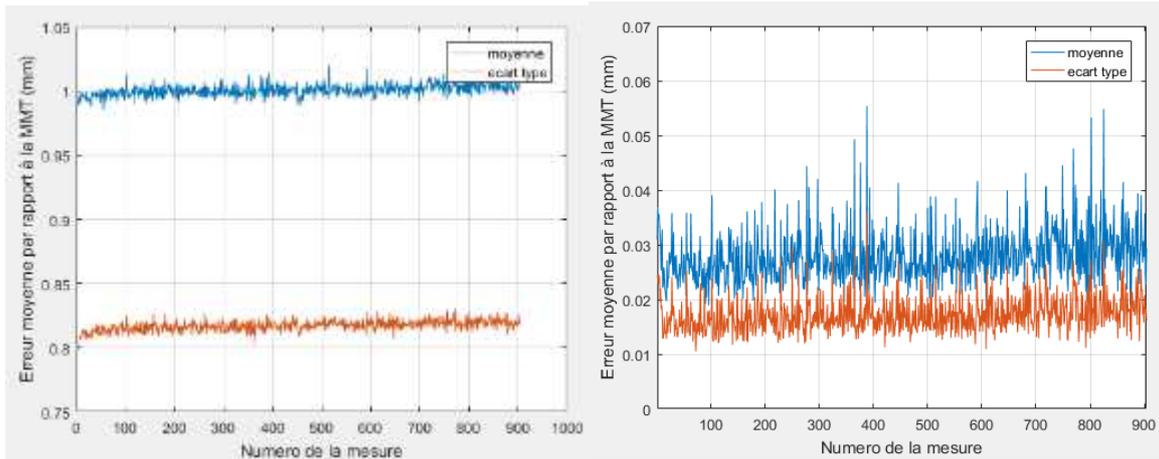

*Figure 7. Evolution of the mean measurement error during 42hours at ambient pressure without correction of the data (left) and with correction of the data (right). The mean error value is indicated in blue, the standard deviation of the value (ie the standard deviation from the reflectors values) is indicated in red.*

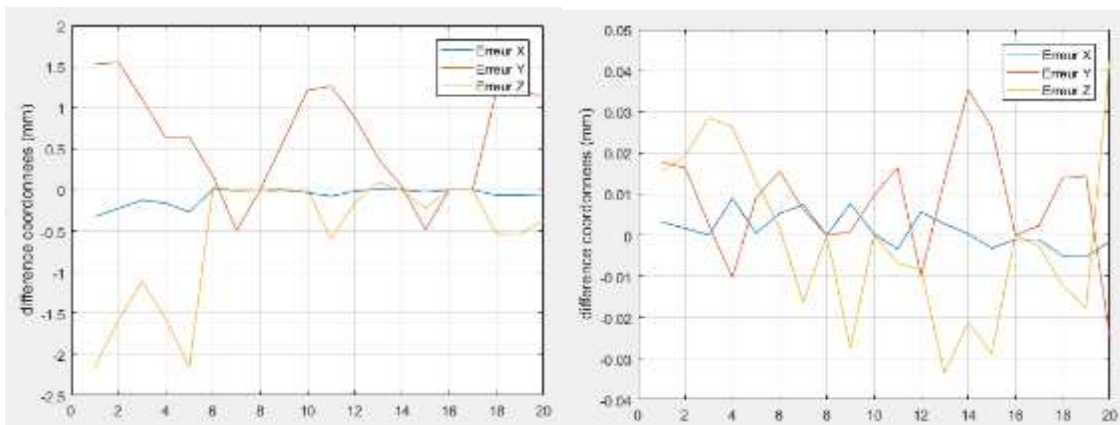

*Figure 8. Error on the X, Y and Z average coordinates over the 900 measurements for each reflector measured without correction of the data (left) and with correction of the data (right) at ambient pressure.*

### 4.3    Measurement repeatability with the window at ambient pressure

In addition to the impact of the window, we have studied the measurement repeatability over time at ambient pressure to validate the measurement repeatability in ERIOS environment. The positions of the reflectors in ERIOS are measured each 2 minutes during 900 measurements to analyze the repeatability (and thus the stability) of the measurement. We study the variation during time of the values measured for each reflector position in X, Y and Z direction i.e. values at moment N – values at moment 0. Figure 9 shows the results obtained at ambient pressure during 42 hours. We can see that:

- The measurement repeatability is much better in the X axis than in the other axis. This is consistent with the conclusion indicated in subsection 3.2. It is explained by the fact that this direction is measured with the distancemeter that is much accurate. The uncertainties on the X measurement is better than +/-15µm;
- The measurement repeatability is worse in Y and Z axis but it is lower than the laser tracker uncertainties estimated in subsection 3.2.  ie < 60µm. We can see that the Z axis measurement is more noisy than the Y axis measurement but both are lower than +/-60µm.

From this measurement and analysis, we can conclude that the laser tracker measurement through the window has the same uncertainties as the measurement of the laser tracker without the window at ambient pressure. No additional terms due to the window are added.

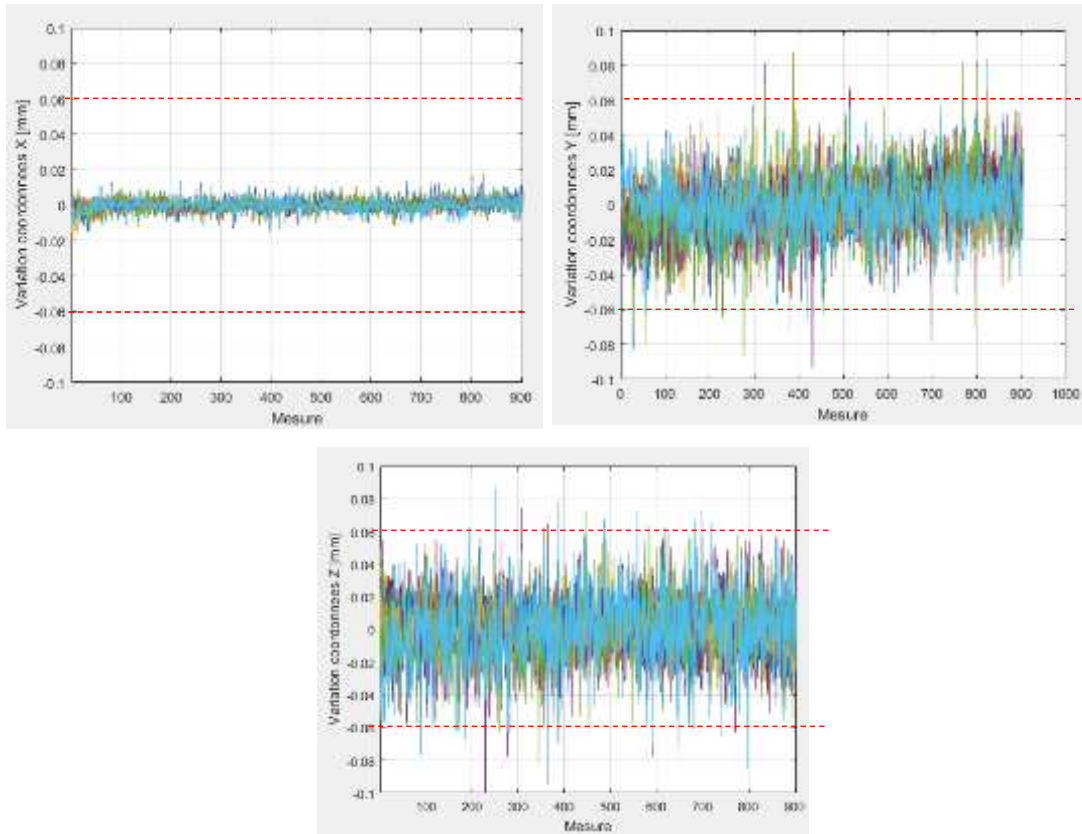

*Figure 9. Variation of the X, Y and Z coordinates of every reflectors measured (20 reflectors) during 42h at ambient pressure. The red curve indicates the laser tracker incertitude calculated in subsection 3.2.*

## 5. MEASUREMENT OF THE LASER TRACKER THROUGH A WINDOW AT VACUUM PRESSURE

The final step and the main goal of the verification plan is the validation of the measurement with the laser tracker at vacuum. This is particularly important for us as we have to validate the correction term to apply to the measurement obtained through the window and at vacuum and to analyze the measurement accuracy reached in such configuration. The results presented hereafter are the results of the test campaign done at LAM in ERIOS vacuum chamber with the mechanical standard in September 2017 according to the test configuration shown in subsection 2.2.

### 5.1    Measurement corrective term in vacuum

The principle of the measurement of the laser tracker through a curved window at vacuum is similar to the case with the window, but in this configuration, the index between the window and the reflectors is not the one of air but of vacuum. This brings a little bit more complexity in the corrective term to be applied to the data. The laser tracker is kept aligned with respect to the curved window but the environment of the measurement changes between the laser tracker and the reflectors. In addition to the window, a large part of the measurement is done in vacuum i.e. with air index equal to 1 instead of air index depending on temperature, pressure and humidity of the air. The laser tracker calculates the air index for each measurement and corrects the data from the air index. When the measurement is done in vacuum, the laser tracker does not "know" that the environment is not the same between the reflectors and itself. We have thus to correct for the "ignorance" of the laser tracker. Based on the analysis of the configuration and the light propagation, a correction term has been proposed in [5]:

- $\rho_{corr} = \rho_{mes} – 13.93mm + (n_a -1)*( \rho_{mes} – 13.93 -Rc)$ where $n_a$ is the air index calculated with pressure, temperature and humidity according to the Edlen formula, Rc is the curvature radius of the window (300mm in our case);
- θ and φ are not corrected. Errors on their values are negligible.

For analysis of the test result in vacuum, we use the data reduction process shown in Figure 6 with the corrective term defined in this section. We will do the same analysis as the one proposed in subsections 4.2 and 4.3.

### 5.2 Measurement error in vacuum

First, we focus on the measurement error due to the vacuum and the validation of the corrective factor defined for measurement in vacuum. The measurement error is the difference between the real value of the reflector position in R_et and the value of the reflector position obtained after correction of the data defined in subsection 5.1. The measurement error is analyzed for a set of 1000 measurements acquired each minute during 48 hours. Results are shown in Figure 10 and Figure 11. The conclusion is similar to the ones from the analysis at ambient pressure: the residual error with respect to the reference is lower than the laser tracker uncertainties. We are correcting the data with the good term and the residual error is included between 20 and 40µm. The order of magnitude of residual is similar to the results from the ambient pressure test.

From this analysis, we can conclude that the correction of the data at vacuum pressure is correct. The measurement error after correction is lower than the laser tracker uncertainties. We have a bias of the data values on the reflectors location in R_et, maybe a little bit higher than at ambient pressure. Figure 11 shows the error on the X, Y and Z coordinates of the reflectors for the mean values calculated on the 1000 measurements acquired. The error on the values is very similar to the errors at ambient pressure: the vacuum adds a bias that we are able to correct with the formula defined in subsection 5.1.

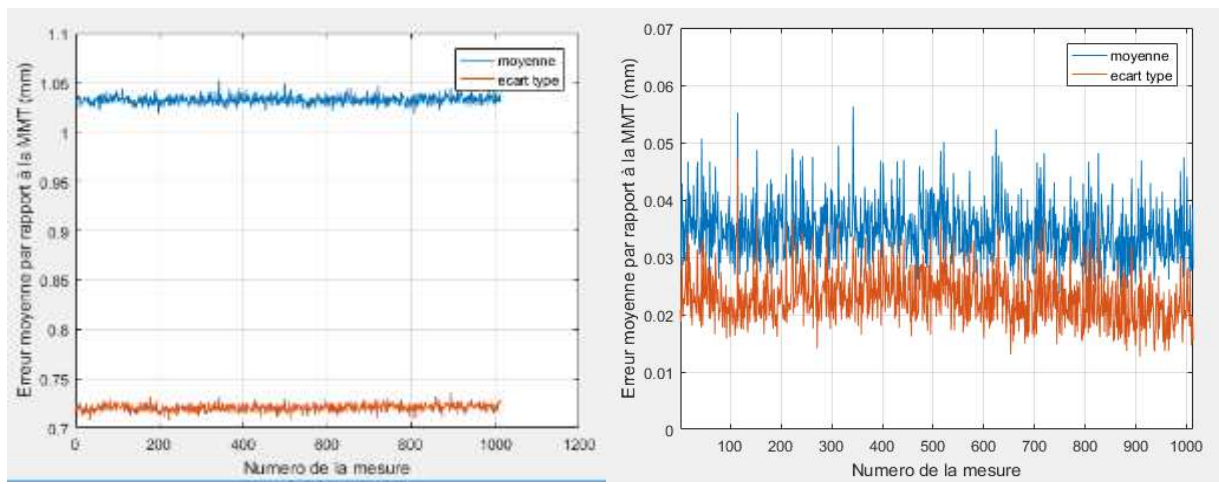

*Figure 10. Evolution of the mean measurement error during 48hours at vacuum pressure without correction of the data (left) and with correction of the data (right). The mean error value is indicated in blue, the standard deviation of the value (ie the standard deviation from the reflectors values) is indicated in red.*

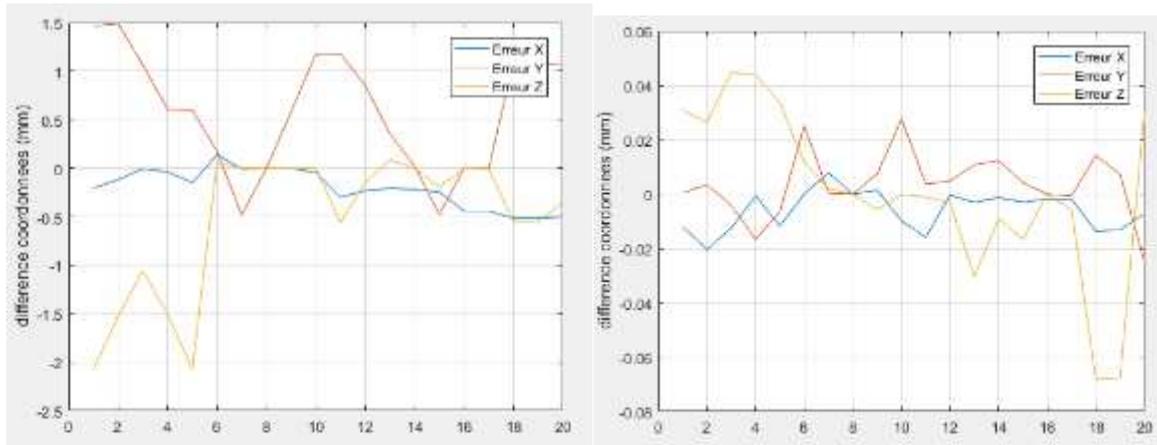

*Figure 11. Error on the X, Y and Z mean coordinates over the 1000 measurements for each reflector measured without correction of the data (left) and with correction of the data (right) at vacuum.*

### 5.3 Measurement repeatability

As for the test at ambient pressure, we analyze the repeatability (and stability) of the measurement. The positions of the reflectors in ERIOS are measured each minute over 1000 measurements (~48h). We study the variation during time of the values measured for each reflector position in X, Y and Z direction i.e. values at moment N – values at moment 0. Figure 12 shows the results obtained at vacuum during 48 hours. We can see that:

- The measurement repeatability is much better in the X axis than in the other axis. This is consistent with the conclusion indicated in the subsection 3.2 and 4.3. Compared to Figure 9, we observed an increase of the standard deviation of the measurement, which is explained by the vacuum condition. In fact, the environment is more noisy when measurement are done in vacuum as several pumps are working and then vibrates in ERIOS environment. The uncertainties on the X measurement is better than +/-15µm;
- The measurement repeatability is worse in Y and Z axis but it is lower than the laser tracker uncertainties estimated in subsection 3.2 i.e. < 60µm. We can see that the Y axis measurement is more noisy than in ambient pressure and similar to results from Z axis. As for X axis, the measurements are in general more noisy due to the ERIOS environment. Both uncertainties are lower than +/-60µm.

From this measurement and analysis, we can conclude that the laser tracker measurement at vacuum has the similar uncertainties as the measurement of the laser tracker without the window and with a window. No additional terms due to the vacuum need to be added but a measurement repeatability of 60µm should be considered.

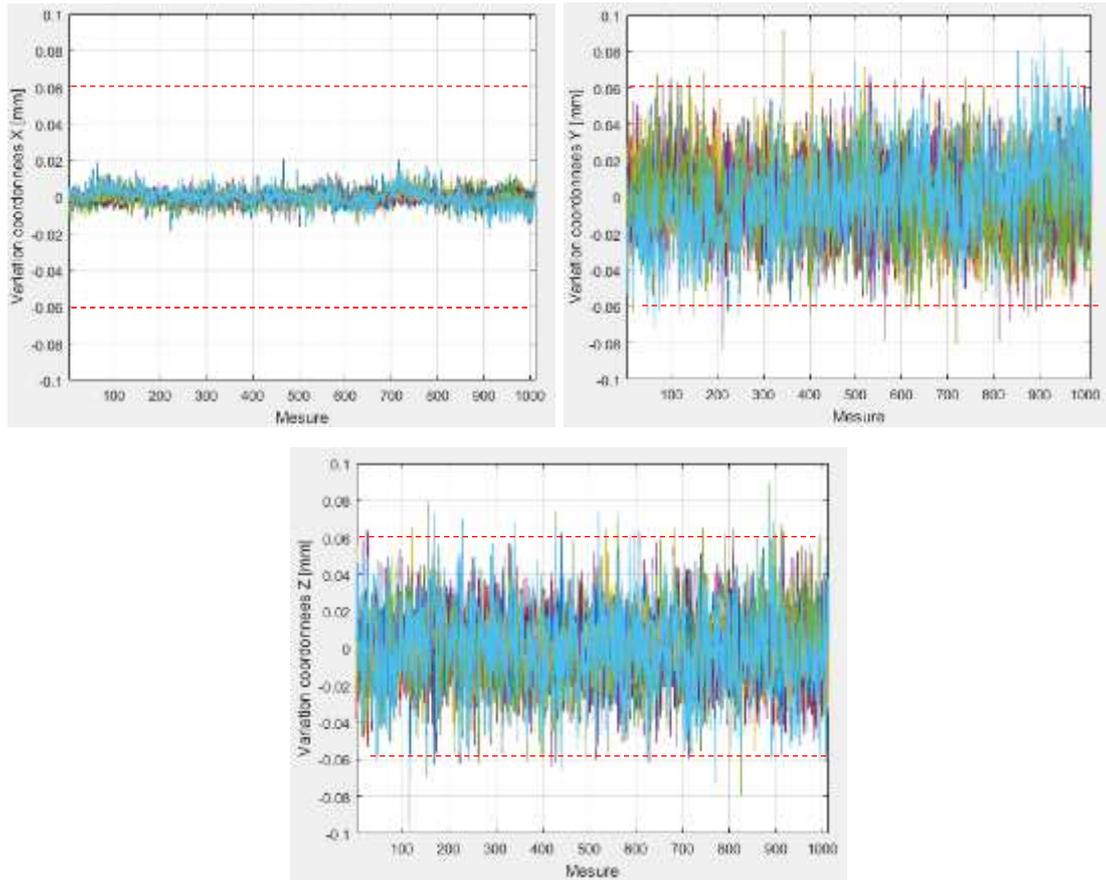

*Figure 12. Variation of the X, Y and Z coordinates of every reflectors measured (20 reflectors) during 48h at vacuum. The red curve indicates the laser tracker incertitude calculated in subsection 3.2.*

## 6. CONCLUSION

We presented in this article the metrology means for the NISP test campaign and we focused on the verification plan of this special device. We showed the test results of the verification campaign. In particular, we identified the corrective factor to be applied to the laser tracker data due to the window and the vacuum pressure. We showed that no additional error term should be applied to the data in such configuration. The study of the measurement repeatability showed that in both configurations the results are lower than the laser tracker uncertainties estimated to 60µm. The next step of the validation plan is to perform a test at cold during the acceptance test of the EUCLID telescope simulator to be ready for NISP test campaign by end of 2018.